%% file: _main.tex
\input{_constants}
\arxiv 

\pdfoutput=1
\documentclass[10pt,twocolumn,letterpaper]{article}
\input{cvpr_header}

\unless\ifarxiv \myexternaldocument{_supplementary} \fi

\newcommand{\myparagraph}[1]{\vspace{0.2em}\noindent\textbf{#1}}

\usepackage{bm}
\usepackage{amsmath}
\usepackage{algorithm}
\usepackage{algpseudocode}

\usepackage{xcolor}
\usepackage{pifont}

\newcommand\footnoteref[1]{\protected@xdef\@thefnmark{\ref{#1}}\@footnotemark}

\begin{document}
\title{UniPhys: Unified Planner and Controller with Diffusion for Flexible Physics-Based Character Control}
\author{\authorBlock}

\twocolumn[{%
\renewcommand\twocolumn[1][]{#1}%
\vspace{-0.5cm}
\maketitle


\begin{center}
  \newcommand{\teaserwidth}{\textwidth}
  \vspace{-0.5cm}
  \centerline{\includegraphics[width=1.0\linewidth]{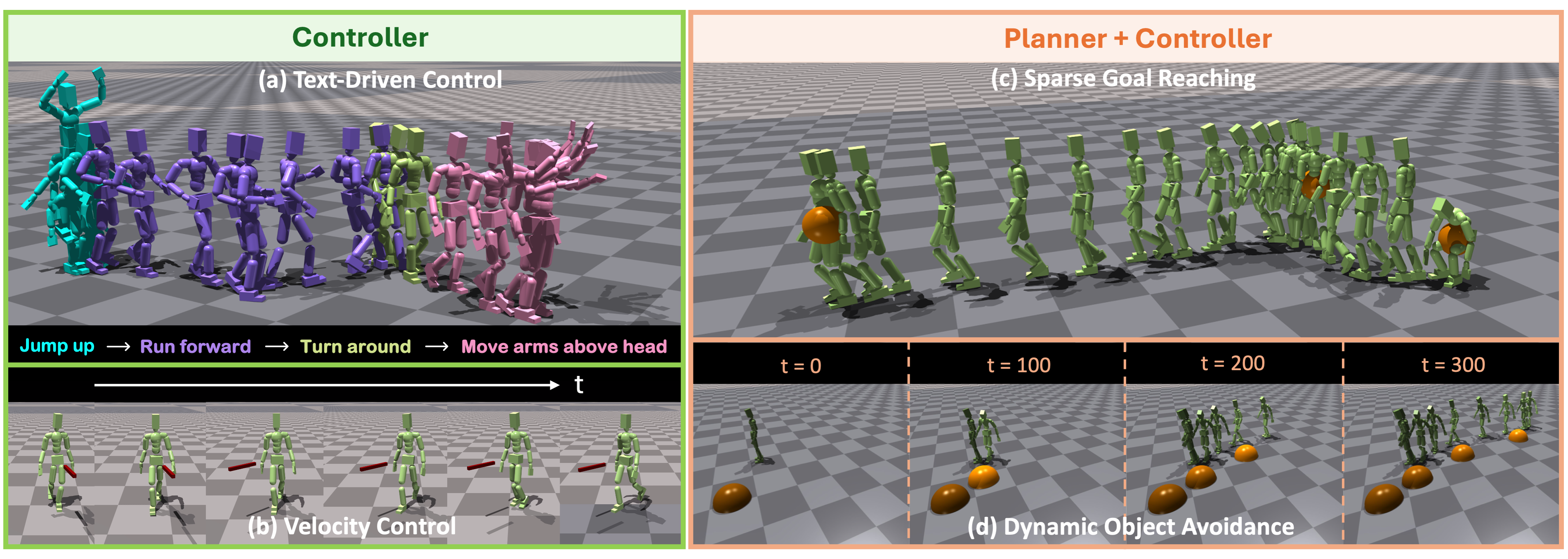}}
    \captionof{figure}{UniPhys is a diffusion-based unified planner and controller for physics-based character control, handling diverse tasks with a single model. We showcase its effectiveness in (a) text-driven control with dynamic language instructions, (b) precise velocity control, (c) sparse goal reaching, and (d) adapting to dynamic environments with moving object avoidance.
    }
    \label{fig:teaser}
\end{center}%
}]

\input{00_abstract}
\input{01_intro}

\input{02_related}

\input{03_method}

\input{04_experiments}

\input{10_conclusion}

\input{references.bbl}


\clearpage
\input{12_appendix}

\end{document}

%% file: _constants.tex
\def\paperID{405}
\def\confName{ICCV}
\def\confYear{2025}

\def\authorBlock{
    Yan Wu\textsuperscript{1} \qquad
    Korrawe Karunratanakul\textsuperscript{1} \qquad
    Zhengyi Luo\textsuperscript{2} \qquad
    Siyu Tang\textsuperscript{1} \qquad
    \\
    \textsuperscript{1} ETH Zurich, \textsuperscript{2} Carnegie Mellon University \\
    {\tt\small \{yan.wu, korrawe.karunratanakul, siyu.tang\}@inf.ethz.ch, zluo2@andrew.cmu.edu}\\
    \small
    \url{https://wuyan01.github.io/uniphys-project/}
}

\newif\ifreview 
\newif\ifarxiv \newcommand{\arxiv}{\arxivtrue}
\newif\ifcamera 
\newif\ifrebuttal 

%% file: cvpr_header.tex
\ifreview \usepackage[review]{cvpr} \fi
\ifarxiv \usepackage[pagenumbers]{cvpr} \fi
\ifrebuttal \usepackage[rebuttal]{cvpr} \fi
\ifcamera \usepackage{cvpr} \fi

\input{_macros}  

\usepackage{xr-hyper}

\makeatletter
\newcommand*{\addFileDependency}[1]{
  \typeout{(#1)}
  \@addtofilelist{#1}
  \IfFileExists{#1}{}{\typeout{No file #1.}}
}

\makeatother
\newcommand*{\myexternaldocument}[1]{
    \externaldocument{#1}
    \addFileDependency{#1.tex}
    \addFileDependency{#1.aux}
}

\definecolor{cvprblue}{rgb}{0.21,0.49,0.74}
\usepackage[pagebackref,breaklinks,colorlinks,allcolors=cvprblue]{hyperref}
\usepackage[capitalize]{cleveref}
\crefname{section}{Sec.}{Secs.}
\crefname{table}{Table}{Tables}
\crefname{figure}{Fig.}{Figs.}

\ifarxiv \crefname{appendix}{App.}{Apps.}
\else \crefname{appendix}{Suppl.}{Suppls.} \fi

%% file: _macros.tex
\usepackage{graphicx}	
\usepackage{amsmath}	
\usepackage{amssymb}	
\usepackage{booktabs}
\usepackage{times}
\usepackage{microtype}
\usepackage{epsfig}
\usepackage{caption}
\usepackage{float}
\usepackage{placeins}
\usepackage{color, colortbl}
\usepackage{stfloats}
\usepackage{enumitem}
\usepackage{tabularx}
\usepackage{xstring}
\usepackage{multirow}
\usepackage{xspace}
\usepackage{url}
\usepackage{subcaption}
\usepackage{xcolor}
\usepackage[hang,flushmargin]{footmisc}

\ifcamera \usepackage[accsupp]{axessibility} \fi

% ...

\ifarxiv  \fi

\newcommand{\R}[1]{{%
    \textbf{%
        \ifstrequal{#1}{1}{\textcolor{red}{R#1}}{%
        \ifstrequal{#1}{2}{\textcolor{blue}{R#1}}{%
        \ifstrequal{#1}{3}{\textcolor{magenta}{R#1}}{%
        \ifstrequal{#1}{4}{\textcolor{teal}{R#1}}{%
                           \textcolor{cyan}{R#1}%
        }}}}%
    }%
}}

%% file: 00_abstract.tex
\begin{abstract}

Generating natural and physically plausible character motion remains challenging, particularly for long-horizon control with diverse guidance signals. 
While prior work combines high-level diffusion-based motion planners with low-level physics controllers, these systems suffer from domain gaps that degrade motion quality and require task-specific fine-tuning.
To tackle this problem, we introduce UniPhys, a diffusion-based behavior cloning framework that unifies motion planning and control into a single model. UniPhys enables flexible, expressive character motion conditioned on multi-modal inputs such as text, trajectories, and goals. To address accumulated prediction errors over long sequences, UniPhys is trained with the Diffusion Forcing paradigm, learning to denoise noisy motion histories and handle discrepancies introduced by the physics simulator. This design allows UniPhys to robustly generate physically plausible, long-horizon motions. 
Through guided sampling, UniPhys generalizes to a wide range of control signals, including unseen ones, without requiring task-specific fine-tuning. 
Experiments show that UniPhys outperforms prior methods in motion naturalness, generalization, and robustness across diverse control tasks.

\end{abstract}

%% file: 01_intro.tex
\section{Introduction}
\label{sec:intro}
Generating natural and physically plausible character motion is essential for graphics, games, animations, and robotics applications.
%
%
Prior research has shown that physics-based character control can be formulated as tracking the reference motion clips from motion capture datasets with Goal-conditioned Reinforcement Learning (RL) in a physics simulator \cite{schulman2015high, duan2016benchmarking, yu2018learning, 2018-TOG-deepMimic, Luo2023PerpetualHC}.
The tracking policy can be distilled with supervised learning into a more versatile controller that can accept multi-modal signals such as text prompt, keyframes, target joint location, etc \cite{tessler2024masked,truong2024pdp,juravsky2024superpadl}.
However, robust multi-task policies for bipedal character control remain challenging, and existing methods are still limited in the naturalness of the generated motion~\cite{tessler2024masked}. They struggle to generalize across diverse control signals~\cite{juravsky2024superpadl} and support only a limited range of motion behaviors~\cite{truong2024pdp}.  
Moreover, when pursuing long-term goals, such as tasks involving complex action sequences or distant objectives, these controllers typically rely on hand-crafted heuristics or manually designed intermediate targets to guide the character’s behavior.

Recent works have shown that physics-based character controllers can benefit from integrating high-level planners~\cite{wang2020unicon, tevet2024closd, rempeluo2023tracepace, Wang2024PacerPlus}, such as diffusion-based generative motion models, which generate intermediate targets to enable controllers to achieve longer-term goals and complex tasks~\cite{rempeluo2023tracepace, tevet2024closd}. These diffusion-based motion planners support various multimodal conditioning signals~\cite{xie2023omnicontrol, wan2024tlcontrol} and arbitrary loss guidances~\cite{song2023loss, karunratanakul2024optimizing}. For example, \cite{tevet2024closd} employs a diffusion model as a kinematic pose planner guiding an RL-based low-level tracking controller in a closed-loop manner.
However, these approaches often result in lower motion quality compared to purely kinematic methods, partially due to not enough tight integration between the motion planner and motion generator. 
This gap also leads the control policy to struggle in accurately tracking planned motions, necessitating fine-tuning for specific tasks and limiting their generalization capabilities~\cite{tevet2024closd}.
To bridge this gap, we introduce UniPhys, a behavior cloning framework that seamlessly integrates both planning and control into a single model, eliminating the domain gap between these components for flexible and human-like character control. UniPhys is a diffusion-based policy model that enables natural and expressive motion generation, allowing control via text or arbitrary guidance signals, akin to guided motion generation in kinematics-based models~\cite{karunratanakul2023guided}.


Our key insight is that the primary challenge for behavior cloning in achieving long-term planning and robust control is the accumulated error at each step of autoregressive prediction. 
By mitigating this compounded error, the resulting model can both plan and control without relying on a low-level RL policy for motion execution. 
To do so, we trained UniPhys following the Diffusion Forcing paradigm \cite{chen2025diffusionforcing}, where the diffusion model learns to denoise sequences with frames with varying noise levels. 
During inference, the model can treat past predictions as slightly noisy to account for error propagation and changes introduced by the physics simulator. This idea is in contrast to typical autoregressive models that assume a clean history. 
This process is illustrated in Fig.~\ref{fig:overview}(b). 

We show that UniPhys can effectively generate physically plausible, long-horizon character motions, conditioned on a range of objectives and guidance signals including those unseen during training. We evaluate UniPhys across various tasks including text-driven control, velocity control, sparse goal-reaching, and dynamic obstacle avoidance. Unlike previous methods, UniPhys produces more natural motions without requiring per-task fine-tuning and is not restricted to a limited set of actions. Our key contributions are:
\begin{itemize}
    \item We introduce UniPhys, a diffusion-based method that unifies the planner and controller for flexible physics-based character control tasks conditioned on arbitrary objectives. It can produce long-horizon, natural, and robust motions that align well with text instruction. 
    \item We propose various guided sampling techniques and task-specific losses suitable for each task. The same model can complete each task without per-task fine-tuning.
    \item We construct, and will release upon publication, a large-scale physics character motion state-action dataset, with frame-level text annotation from BABEL~\cite{BABEL:CVPR:2021}, that can be used for imitation learning.
\end{itemize}

%% file: 02_related.tex
\section{Related Work}
\label{sec:related}
\myparagraph{Human motion synthesis.}
Significant efforts have been devoted to capturing human motion and annotating textual descriptions for motion sequences~\cite{AMASS:ICCV:2019, BABEL:CVPR:2021, Guo_2022_CVPR, plappert2016kit}. Building on these rich resources, kinematics-based human motion generation has achieved remarkable progress in synthesizing natural movements using diverse conditional inputs such as text~\cite{Guo_2022_CVPR, kim2022flame, petrovich2022temos}, music~\cite{lee2019dancing, li2022danceformer, tseng2023edge, alexanderson2023listen}, and other modalities~\cite{huang2023diffusion, petrovich2021action, Rempe2021-hp, Shafir2024-qh, karunratanakul2023guided}. The emergence of diffusion models \cite{ho2020denoising, song2019generative, song2020score} has further enhanced the expressiveness of these approaches, enabling finer control over motion synthesis~\cite{tevet2023human, zhang2024motiondiffuse, chen2023executing}. 
However, such methods often produce physically implausible artifacts such as foot sliding and floating due to the lack of physics constraints. In contrast, physics-based character control inherently enforces realism and plausibility by grounding motion in physical simulators but struggles to match the expressiveness, diversity, and scalability of kinematics-based methods yet~\cite{tessler2024masked, truong2024pdp, juravsky2024superpadl, yao2024moconvq}. Bridging these paradigms by combining the plausibility of physics with the expressiveness of data-driven kinematics remains an open challenge.


\myparagraph{Physics-based character animation.} 
Achieving natural human and animal character control has been a long-standing challenge in computer animation \cite{lee2002interactive, lee2010motion, safonova2007construction, treuille2007near, levine2012continuous, zhang2018mode, holden2017phase, ling2020character}.
Recent advances in physics-based character control focus on replicating large-scale MoCap datasets using reinforcement learning (RL) \cite{2018-TOG-deepMimic, wang2020unicon, won2020scalable, Luo2023PerpetualHC, ScaDiver, wagener2022mocapact} and imitation learning \cite{peng2021amp, peng2022ase, tessler2023calm, tessler2024masked}. A key approach involves learning motion priors from MoCap data for downstream control. AMP \cite{peng2021amp} trains a physics-based control policy using an adversarial discriminator for motion realism, but requires separate policy training for each task. Subsequent methods like ASE \cite{peng2022ase}, CALM \cite{tessler2023calm}, ControlVAE \cite{ControlVAE}, and PULSE \cite{luo2024universal} aim to distill more generalized motion priors from tracking policy, however, they still require task-specific controllers training. MaskedMimic \cite{tessler2024masked} improves this by learning a multi-task controller using a masked conditional VAE, but struggles to generalize beyond predefined control signals.


Another research area explores text-driven control policies~\cite{tessler2023calm, juravsky2022padl, juravsky2024superpadl, truong2024pdp}. SuperPADL \cite{juravsky2024superpadl} uses multi-stage reinforcement learning and behavior cloning to create a versatile text-driven policy. PDP \cite{truong2024pdp} employs diffusion models for a multi-modal text-driven policy via behavior cloning, reducing errors by adding noise during data collection. Despite their promising results, these methods lack controllability, restricting their ability to generate behaviors according to novel guidances. 
Currently, physics-based text-driven policies still trail kinematics-based approaches in motion diversity, expressiveness, and scalability due to challenges in distilling controllable, multi-modal, and robust policies.


To address this gap, hierarchical frameworks are gaining popularity \cite{tevet2024closd, hansen2024hierarchical, rempeluo2023tracepace, Wang2024PacerPlus}. These methods divide control into two stages: (1) a high-level planner generating waypoints \cite{rempeluo2023tracepace}, joint trajectories \cite{tevet2024closd}, or partial-body targets \cite{hansen2024hierarchical}, and (2) a low-level RL controller tracking these plans. For instance, CLoSD \cite{tevet2024closd} integrates a diffusion-based kinematic planner with an RL tracker. However, misalignment between kinematic plans and physical constraints can cause unnatural artifacts like jittering and foot skating, necessitating additional task-specific controller fine-tuning \cite{tevet2024closd, hansen2024hierarchical}.

\myparagraph{Diffusion model for planning and control.}
Diffusion models are effective for both planning \cite{janner2022diffuser, chen2025diffusionforcing,carvalho2023motion} and control \cite{chi2023diffusionpolicy} due to their capacity to handle multi-modal distributions and incorporate conditioning signals like text and goals. In robotics, these models can map observations into actions for tasks such as manipulation and navigation \cite{huang2024diffuseloco, chi2023diffusionpolicy,ma2024hierarchical,sridhar2024nomad}. However, their application in high-dimensional control systems like physics-based characters remains underexplored. 
Critically, existing frameworks for character control often separate planning and control, with diffusion models producing either high-level plans or low-level actions. In this work, we integrate planning and control for physics-based characters into a single diffusion framework, thereby, eliminating the hierarchical discrepancies. Our method optimizes both kinematic realism and physical plausibility, achieving expressive, text-aligned motions while maintaining dynamic feasibility.


%% file: 03_method.tex

\section{Preliminary}
\label{sec:preliminary}

\myparagraph{Physics Simulation Setup.} We control a SMPL-like~\cite{SMPL:2015} physics-based character in the Isaac Gym simulator~\cite{liang2018gpu}, featuring 24 rotational joints, with 23 actuated, excluding the pelvis. Each actuated joint uses proportional-derivative (PD) control, and the action $\mathbf{a_t} \in \mathbb{R}^{J\times3}$ specifies the target joint positions. The simulator provides the character state $\mathbf{s_t}$ and calculates the dynamic transition $\mathbf{s}_{t+1} = \mathcal{SIM}(\mathbf{s}_t, \mathbf{a}_t)$.



\myparagraph{Physics-based character tracking policy.} 
Previous work has used goal-conditioned reinforcement learning to replicate MoCap datasets in a physics-based simulator for tracking reference motions~\cite{2018-TOG-deepMimic, Luo2023PerpetualHC}. PHC~\cite{Luo2023PerpetualHC} effectively tracks the entire AMASS dataset \cite{AMASS:ICCV:2019} with a single policy, $\mathbf{a}_t = \boldsymbol{\pi}_{PHC}(\mathbf{s}_t, \mathbf{s}_t^g)$\footnote{\label{note:phc}PHC and PULSE tracking policy takes the proprioception state $\mathbf{s}_t^p$ as input, normalizing the global state $\mathbf{s}_t$ to the local body frame. For notation simplicity, we omit this step and directly use $\mathbf{s}_t$ to represent the input state.}, where $\mathbf{s}_t^g$ represents the next-frame goal states. This policy is optimized with PPO~\cite{Schulman2017ProximalPO}, using a reward function that aligns the induced next state $\mathbf{s}_{t+1} = \mathcal{SIM}(\mathbf{s}_t, \mathbf{a}_t)$ with the goal state $\mathbf{s}_t^g$.

\begin{figure}
    \centering
    \includegraphics[width=0.95\linewidth]{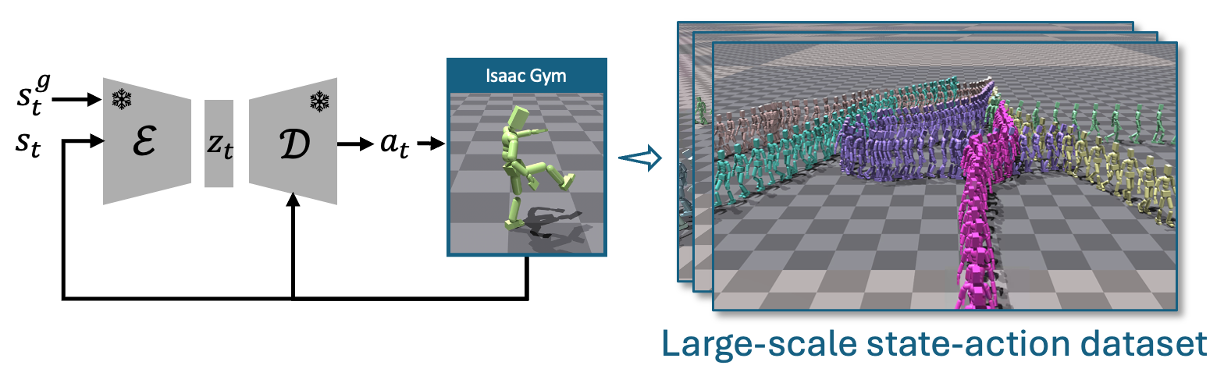}
    \caption{We construct a large-scale paired state-action dataset by tracking MoCap dataset with PULSE tracking policy~\cite{luo2024universal}.}
    \label{fig:dataset-curation}
\end{figure}

\myparagraph{Physics-based motion latent space.}
Expanding on the tracking policy, PULSE~\cite{luo2024universal} distills the PHC tracking policy into a physics-based latent motion space with conditional variation autoencoder (cVAE) for generative control. The encoder maps $\mathbf{s}_t$ and $\mathbf{s}_t^g$ to a latent embedding $\mathbf{z}_t$, while the decoder reconstructs the action needed to track $\mathbf{s}_t^g$, as presented in Fig.~\ref{fig:dataset-curation}. Mathematically, $\mathbf{z}_t \sim \mathbf{\mathcal{E}}(\mathbf{z}_t|\mathbf{s}_t, \mathbf{s}_t^{g})$, $\mathbf{a}_t = \mathbf{\mathcal{D}}(\mathbf{s}_t, \mathbf{z}_t)$. Training is performed via online distillation, with supervision signals from the tracking policy $\boldsymbol{\pi}_{PHC}(\mathbf{s}_t, \mathbf{s}_t^g)$. 

Prior works have demonstrated that the distilled latent space provides a well-regularized action space~\cite{luo2024universal, ControlVAE, PhysicsVAE, yao2024moconvq, peng2022ase}, allowing efficient learning of downstream tasks via reinforcement learning: $\mathbf{z}_t = \boldsymbol{\pi}_{task}(\mathbf{o}_t, \mathbf{g}_t^{task})$, where $\mathbf{o}_t$ is the current observation and $\mathbf{g}_t^{task}$ is the task goal. 
We observe that, because the latent embedding $\mathbf{z}_t$ captures the dynamic transition between consecutive frames, it can also serve as a generalized action representation for all tasks. Thus, we employ this method for dataset curation and use $\mathbf{z}_t$ as the action representation for our model to predict.

\section{UniPhys: Unified Planner and Controller}
We aim to create a unified planning-control framework that addresses inconsistencies in the two-stage planner-controller paradigms while enabling zero-shot generalization across various control tasks. 
To this end, we introduce a diffusion-based generative behavior model that simultaneously learns action distributions and dynamic state transitions.
First, we explain how we curate a large-scale offline dataset of physics-based character motions suitable for behavior cloning training in Sec.~\ref{sec:dataset}.
The architecture and training of our model are discussed in Sec.\ref{sec:Behavior-Generative-Model}.
Our guided control framework, which allows for flexible and adaptable task control during inference, is introduced in Sec.\ref{sec:sampling}.
Finally, we demonstrate the versatility and effectiveness of our framework across multiple applications in Sec.~\ref{sec:application}.

\subsection{Dataset Curation}
\label{sec:dataset}
Large-scale motion capture datasets, such as AMASS~\cite{AMASS:ICCV:2019}, offer extensive human motion data, while text-annotated subsets like HumanML3D~\cite{Guo_2022_CVPR} and BABEL~\cite{BABEL:CVPR:2021} offer complementary semantic information. However, there are limited publicly available datasets with state-action sequences and text descriptions suitable for learning physics-based control policies.
To address this, we created a large-scale state-action dataset to enable policy learning via behavior cloning. We tracked motions from the AMASS dataset using PULSE tracking policy~\cite{luo2024universal}, storing successfully tracked sequences with paired state-action data and latent embeddings for behavior cloning, i.e., ($\mathbf{s}_t, \mathbf{a}_t, \mathbf{z}_t$).
Additionally, we added frame-level text annotations from the BABEL dataset to enrich the dataset with semantic information. These detailed atomic action labels enable learning a variety of text-driven atomic skills and allow for the flexible composition of skills to perform complex tasks.
Ultimately, we compile a paired text-state-action dataset with 4,875 sequences from the BABEL training set, totaling 15.7 hours of motion. We will release this dataset publicly to facilitate research in imitation learning for physics-based character control.
Implementation details are available in the Supp. Mat.



\begin{figure*}
    \centering
    \includegraphics[width=0.98\linewidth]{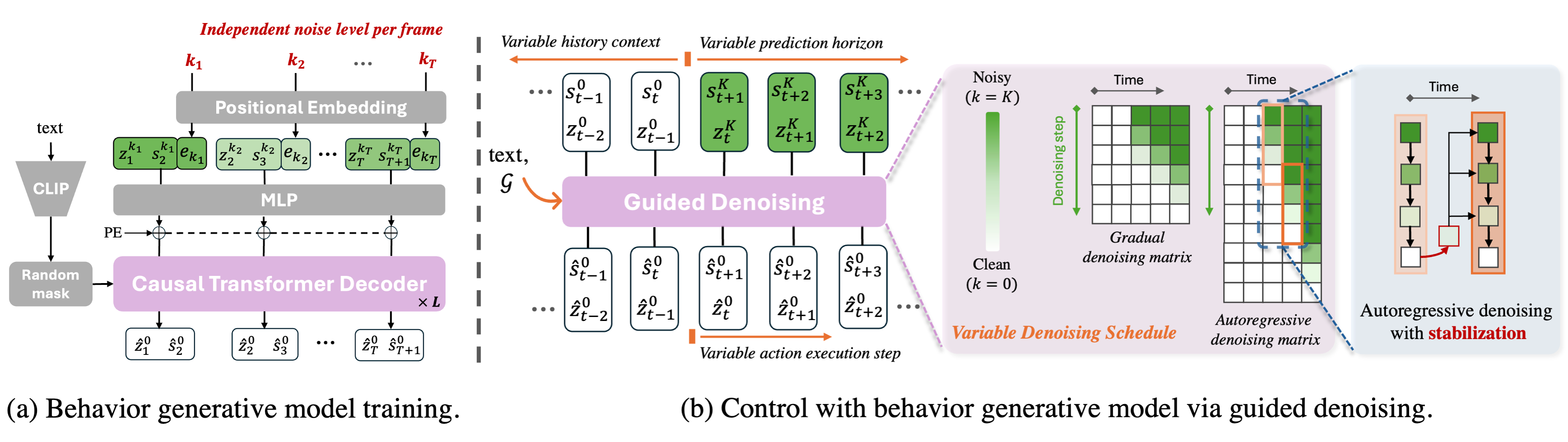}
    \caption{Framework overview. (a) The model takes a behavior sequence of length T as input and is conditioned on the clip-based text embedding. At training time, each frame is corrupted with different noise levels, and the model learns to predict the clean behavior sequence. (b) At test time, guided denoising with task-specific guidance enables flexible multi-task control. We highlight the \textcolor{orange}{flexibility} in different test-time denoising conditions and configurations, and the \textcolor{red}{stabilization} trick that promotes stable long-horizon autoregressive control.}
    \label{fig:overview}
\end{figure*}

\subsection{Diffusion-based Behavior Generative Model}
\label{sec:Behavior-Generative-Model}
Using the paired text-state-action dataset as expert demonstrations, we introduce a diffusion-based generative model that unifies planning and control. Our method jointly models state and action distributions conditioned on text to predict future state-action pairs.
We follow the Diffusion Forcing \cite{chen2025diffusionforcing} training paradigm by applying varying noise levels to each frame, unlike typical motion diffusion models that use a uniform noise level across all frames.
Our unified model offers three core capabilities: (1) end-to-end control driven by high-level text instructions; (2) precise state-space control via gradient-based guidance during the diffusion denoising process; and (3) long-horizon planning by simultaneously predicting future states and actions.




\myparagraph{Behavior representation.} 
We define the behavior sequences $\mathbf{X} = \mathbf{x}_{1:T}$, where $\mathbf{x}_t=(\mathbf{s^c}_{t}, \mathbf{z}_{t})$,
to include the canonicalized state sequences $\mathbf{s^c}_{1:T}$ and latent action embedding sequences $\mathbf{z}_{1:T}$. Instead of directly modeling the high-dimensional action space $\mathbf{a}_{t}$, we leverage the well-regularized latent action representation $\mathbf{z}_t \in \mathbb{R}^{32}$ encoded by the PULSE encoder to facilitate efficient action distribution learning. For state representation, we canonicalize the state sequences as $\mathbf{S^c} = (\mathbf{r^c}_{1:T}, \mathbf{p^c}_{1:T}, \mathbf{{v}^c}_{1:T}, \mathbf{q}_{1:T}, \mathbf{w}_{1:T})$, which includes: (1) global root trajectory $\mathbf{r^c}_{1:T} = (\mathbf{\gamma}, \mathbf{\phi}, \dot{\mathbf{\gamma}}, \dot{\mathbf{\phi}})_{1:T}$ canonicalized to the first-frame coordinate system, consisting of root position $\mathbf{\gamma}_t \in \mathbb{R}^3$, orientation $\mathbf{\phi}_t \in \mathbb{R}^6$, linear velocity $\dot{\mathbf{\gamma}}_t \in \mathbb{R}^3$ and angular velocity $\dot{\mathbf{\phi}}_t \in \mathbb{R}^3$. The canonicalized root trajectory always starts from the origin and the first frame faces the y+ axis; (2) local joint features, which are canonicalized to per-frame local coordinate frames, include local joint positions $\mathbf{p^c}_t \in \mathbb{R}^{J\times3}$, velocities $\mathbf{v^c}_t \in \mathbb{R}^{J\times3}$, joint rotation $\mathbf{q}_t \in \mathbb{R}^{J\times6}$, and angular velocity $\mathbf{w}_t \in \mathbb{R}^{J\times3}$. The per-frame local coordinate system is set at the pelvis joint projected on the ground. For rotation, we all adopt the 6D rotation representation. In the following, we omit the canonicalization subscription for notation simplicity, and unless explicitly specified, $\mathbf{s}_t$ and $\mathbf{S}$ indicate canonicalized state and state sequence respectively.

\myparagraph{Training: independent noise injection per frame.} 
As shown in Fig.~\ref{fig:overview}(a), the behavior generative model takes a behavior sequence of length $T$ as both input and output. Each frame represents a single token formed by concatenating the latent action $\mathbf{z}_t$ and the induced next state $\mathbf{s}_{t+1}$, resulting in a 398-dimensional feature per frame. The noise levels can be \textit{independent} for each frame, and the noise level embedding is incorporated into the per-frame feature. We use a causal transformer decoder as the backbone, conditioning the output on a CLIP-based~\cite{radford2021learning} text embedding.
At training time, instead of applying uniform noise level across the entire sequence, we apply \textit{independent and random noise to each token} in the sequence. 

At each training step, the sequence $\mathbf{X}^0$ is corrupted to $\mathbf{X^k}=(\mathbf{x}_1^{k_1}, \mathbf{x}_2^{k_2}, \cdots \mathbf{x}_T^{k_T})$,
    where $\mathbf{x}_t^{k_t} = \sqrt{\Bar{\alpha}_t}\mathbf{x}_t^0 + \sqrt{1-\Bar{\alpha}_t}\mathbf{\epsilon}^{k_t}$, and $\mathbf{\epsilon}^k \sim \mathcal{N}(0, \mathbf{I})$, following the forward diffusion process, and per-frame random noise levels $\mathbf{k}=k_{1:T} \in [K]^T$ are independently randomly sampled. The model is parameterized as $\mathbf{\mathcal{M}}_{\theta}(\mathbf{X^k}, \mathbf{k}, \mathbf{c})$ to predict the clean behavior sequence, where $\mathbf{c}$ is the text embedding, and the training loss is given by:
\begin{align}
\mathcal{L}(\theta) = \mathbb{E}_{\mathbf{k}, \mathbf{X}^0}[||\mathbf{X}^0 - \mathbf{\mathcal{M}}_\theta(\mathbf{X^k}, \mathbf{k}, \mathbf{c})||^2]
\end{align}
More details are discussed in \cite{chen2025diffusionforcing} and in Supp. Mat.

\begin{algorithm}[]
\caption{Test-time control with guided denoising.}
\begin{algorithmic}[1]
\Require Behavior generative model $\mathcal{M}_\theta$ (T frames), 
\Statex \hspace{0.8cm} PULSE action decoder $\mathcal{D}$,
\Statex \hspace{0.8cm} Physics simulator $\mathcal{SIM}$ at simulation step $t$.
\Statex \hspace{-0.7cm} \textbf{Optional input:} Text instruction $\mathbf{c}$, Guidance loss $\mathcal{G}(\cdot)$
\Statex \hspace{-0.7cm} \textbf{Task-specific config.:} History motion $\mathbf{x}_{t-h:t}$ with length $h$,
\Statex \hspace{2.5cm} Prediction horizon $H$, 
\Statex \hspace{2.5cm} Action execution step $T_a$,
\Statex \hspace{2.5cm} Denoising schedule.
\Statex \hspace{-0.7cm} \textbf{Hyperparam.:} stabilization noise level $n$, monte-carlo sample number $N$ (only for loss-based guidance).

\State Initialize $\mathbf{x}_{t+1}, \ldots, \mathbf{x}_{t+H} \sim \mathcal{N}(0, \sigma^2 \mathbf{I})$

\label{alg:inference}
\State Input window $\mathbf{X}=\mathbf{x}_{t-h+1:t-h+T}$
\State $\mathcal{K} \in \mathbb{R}^{M \times T} \gets$ DenoisingMatrix($T, h, H$) \Comment{Fig~\ref{fig:overview}(b)}
\For{row $m = M - 1, \ldots, 0$}
    \State $k$ ← ReplaceZeros($\mathcal{K}_m, n$)   \Comment{Stabilization trick}
    \State $\mathbf{X} = \mathcal{M}_\theta(\mathbf{X}, k)$
    \State $\mathbf{X} \gets$ CFG($\mathbf{X}, c$) with Eq.~\ref{eq:classifier-free}
    \State $\mathbf{X} \gets$ MCG($\mathbf{X}, \mathcal{G}(\mathbf{X})$) with Eq.~\ref{eq:classifier-based}, \ref{eq:MCG}
\EndFor
\For{$i = 1, \ldots, T_{a}$}
    \State $a_{t+i} = \mathcal{D}(\hat{s}_{t+i}, \hat{z}_{t+i})$
\EndFor
\State $s_{t+1:t+T_a+1} \gets \mathcal{SIM}(s_{t}, a_{t:t+T_a})$ 
\end{algorithmic}
\end{algorithm}

\subsection{Guided Behavior Synthesis for Flexible Control}
\label{sec:sampling}
We utilize the diffusion-based behavior model for flexible multi-task control, producing sequential actions through guided denoising.
Overall, our guided denoising-based control framework follows a receding horizon strategy with autoregressive behavior synthesis:
The model is conditioned on past behaviors to iteratively denoise future action tokens. The denoised action tokens are then decoded into executable actions. After executing the predicted action sequence in the simulator, the context window shifts forward, and the cycle repeats, enabling long-horizon rollouts that can also dynamically adapt to task and environmental changes.

The inherent flexibility of diffusion models allows the denoising process to be guided by text prompts for high-level intent and state-based objectives for fine-grained state-space control. Additionally, our per-frame noise injection strategy enables the model to handle \textit{flexible noise configurations} during inference. Leveraging these capabilities, we explore various test-time configurations to optimize performance across different tasks. The inference framework is presented in Fig.~\ref{fig:overview}(b) and Alg.~\ref{alg:inference}. We summarize the key features our model supports at test time as follows:

\myparagraph{Text-conditioned sampling.}
By training the model in a classifier-free manner to condition the generation on text descriptions, we can generate text-driven action sequences with classifier-free guidance (CFG) \cite{ho2022classifier}.

\begin{align}
\small
\hat{\mathbf{X}}^0_c = \mathcal{M}_\theta(\mathbf{X^k}, \mathbf{k}, \emptyset) + \lambda_c(\mathcal{M}_\theta(\mathbf{X^k}, \mathbf{k}, \mathbf{c}) - \mathcal{M}_\theta(\mathbf{X^k}, \mathbf{k}, \emptyset))
\label{eq:classifier-free}
\end{align}
where $\lambda_c$ controls the guidance strength.

\myparagraph{Task-specific loss-guided sampling.} 
Using guided diffusion \cite{dhariwal2021diffusion}, the denoising trajectory can be adjusted according to task-specific objectives.
For each task, we define a loss function $\mathcal{G}(\mathbf{X})$. Then, the denoising process is guided by its gradient toward desired outcomes:
\begin{align}
\hat{\mathbf{X}}_l^0 = \mathcal{M}_\theta(\mathbf{X^k}, \mathbf{k}, \mathbf{c}) - \lambda_l \nabla_{\mathbf{X^k}}\mathcal{G}(\hat{\mathbf{X}}^0),
\label{eq:classifier-based}
\end{align}
where $\lambda_l$ controls the guidance strength. 

In practice, we employ Monte-Carlo Guidance (MCG) \cite{song2023loss} to estimate the gradient from multiple samples. MCG provides a smoother gradient estimate with reduced variance, promoting stable optimization during the denoising process,
\begin{align}
    \nabla'\mathcal{G}(\hat{\mathbf{X}}^0) = \frac{1}{N}\sum_{i=1}^N\nabla_{\mathbf{X^k}}\mathcal{G}(\hat{\mathbf{X}}^0_{(i)})
    \label{eq:MCG}
\end{align}
where $N$ is the number of samples. 


In addition, with the ability to accept different per-frame noise levels during inference, we further explore:

\myparagraph{Variable context length and prediction horizon}.
With the receding horizon strategy, our method enables the generation of both short-horizon actions for reactive control and long-term planning for far-away objectives by changing only the context length and prediction horizon.

\myparagraph{Flexible denoising schedule.} Going beyond the commonly used full-sequence diffusion denoising, we explore two additional strategies: (1) an autoregressive denoising schedule, which denoises the sequence sequentially; (2) a gradual denoising process, which prioritizes denoising near-future frames while preserving uncertainty in distant ones. The sketch diagrams for different denoising schedules are shown in Fig.~\ref{fig:overview}(c). We observe that, compared to full-sequence denoising, autoregressive denoising produces a more stable roll-out, while gradual denoising further improves the performance in long-horizon planning tasks.

\myparagraph{Long-horizon rollouts with stabilization.} 
In behavior cloning, compounding error occurs when small prediction errors accumulate during long-horizon rollouts, leading to distribution drift and unreliable control in out-of-distribution states. To mitigate this, we set the noise indicator of fully denoised frames to be greater than zero, $k>0$, during denoising. This prevents the model from overconfidently treating previous predictions as error-free. Importantly, we only adjust the noise indicator $k$ to signal that previous states are slightly noisy, without adding noise to the state-action predictions. For an overview, see Fig \ref{fig:overview}(b), and for the formal formulation of this stabilization trick, refer to Alg. \ref{alg:inference} line 5. This technique enhances robustness against distributional shifts and improves long-term rollout stability.

By integrating these features, our model can adapt to various tasks, from high-level language-conditioned control and detailed state-space manipulation to long-horizon planning. Flexible test-time denoising configurations allow dynamic adjustment of parameters, such as noise schedules and planning horizons, to meet specific task requirements.

\subsection{Applications}
\label{sec:application}
We demonstrate the versatility of our model on multiple applications across various control levels, including interactive text-driven motion control, sparse goal reaching, velocity control, and dynamic obstacle avoidance. These applications highlight the model's capability to manage both high-level text-driven control and precise motion adjustments while adapting to real-time environmental changes.



\myparagraph{Interactive text-driven controller.} Utilizing fine-grained frame-level annotations from BABEL, our model learns text-aligned motion policies for diverse atomic skills, enabling real-time interactive text-driven control. This allows for on-the-fly text instruction changes and smooth skill transitions.

For interactive text-driven control, we denoise a small portion of the future trajectory, $H=8$ frames. After executing this segment in the simulator, we update the history and repeat the denoising process. This autoregressive rollout mechanism lets the model quickly adapt to changing instructions. To enhance long-term stability, we integrate autoregressive denoising with the stabilization technique, improving rollout robustness and skill transition smoothness.



\myparagraph{Sparse goal reaching.} Using loss-based guidance, our model enables joint position control, crucial for planning and sparse goal-reaching. For this task, We predict with longer future horizon $H=28$ but execute only the first few frames, $T_a=8$, of denoised actions to maintain robust control and adaptability to environmental changes. A gradual denoising schedule is used alongside stabilization, focusing on refining near-future predictions while keeping the distant future uncertain. 
This setup is suitable for long-horizon tasks and can be combined with either loss-based guidance or high-level text instructions to specify different motion styles for actions such as \textit{walking, jogging, running, and sitting}.

To facilitate goal-reaching, we designed a loss function that encourages predicted joint positions to be close to the target. Additionally, an orientation loss is included to encourage the character to face the goal, expediting goal achievement. 



\myparagraph{Velocity control.} Our model can effectively regulate velocity and produce a smooth transition when the target velocity changes. To accomplish this, we designed a loss function to guide predicted velocity toward the desired speed and direction. 
Although long-horizon planning is not crucial for this task, we observe that longer-horizon predictions improve stability and smoothness during transitions when target velocity directions change.
 


\myparagraph{Dynamic object avoidance.} 
Using autoregressive rollouts, our model can also adapt effectively to dynamic environments. We demonstrate this capability in a dynamic obstacle avoidance task, where the character must react to evade a pursuing object. A simple smooth signed distance function (SDF) loss is used to encourage the character to steer away from the obstacle. 
We observe that reducing action execution interval $T_a$ enhances responsiveness to dynamic changes at the cost of efficiency.



Table~\ref{tab:task-config} summarizes different test-time denoising configurations for each application. We provide detailed loss designs in the Supp. Mat.


\begin{table}[]
    \centering
    \footnotesize
    \begin{tabular}{c|c|c|c|c}
    \toprule
         & h & H & $T_a$ & Denoising schedule \\
        \midrule
        Text-driven control & 4 & 8 & 8 & Autoregressive \\
        Goal reaching & 4 & 28 & 8 & Gradual \\
        Speed Control & 4 & 28 & 8 & Gradual \\
        Obstacle avoidance & 4 & 28 & 8 & Gradual \\
        \bottomrule
    \end{tabular}
    \caption{Denoising configurations for different applications, including context frames (h), prediction horizons (H), action execution steps ($T_a$), and the employed denoising schedule.}
    \label{tab:task-config}
\end{table}

%% file: 04_experiments.tex
\section{Experiments}


\label{sec:experiments}

\myparagraph{Baselines.} We compare our method against two state-of-the-art physics-based character multi-task controllers: (1) MaskedMimic \cite{tessler2024masked}, which uses a masked conditional variational autoencoder (cVAE) to distill a multi-task controller from a tracking policy, conditioned on predefined control signals. During training, some control conditions are randomly masked, allowing flexible test-time conditioning control, though it lacks planning capability and cannot generalize to unseen signals; (2) CLoSD \cite{tevet2024closd}, a two-stage framework where a diffusion-based planner generates text- and goal-conditioned kinematic motions, followed by an RL-based tracking controller that tracks the planned kinematic motion. As the planning and control models are separate, 
the control model needs to be fine-tuned on a set of predefined tasks to handle errors induced by the kinematic planner.



\myparagraph{Evaluation metric.} Following PhysDiff~\cite{yuan2023physdiff} and CLoSD~\cite{tevet2024closd}, we assess the physical plausibility of the motion using foot-floating metrics. We do not report foot skating, as it is negligible across all physics-based methods. Additionally, we compute the motion jerk to evaluate the smoothness of the motion. For task-specific evaluations, we introduce additional metrics tailored to each application.


\begin{table*}[t]
    \centering
    \footnotesize
    \renewcommand{\arraystretch}{1.2} 
    \setlength{\tabcolsep}{6pt} 
    \begin{tabular}{lcc|ccc|cc}
        \toprule
        & \multicolumn{2}{c}{\textbf{Text-to-motion precision}}
        & \multicolumn{3}{c}{\textbf{Motion quality user study (Score: 1-5)}} & \multicolumn{2}{c}{\textbf{Physics-based metrics}} \\
        \cmidrule(lr){2-3} \cmidrule(lr){4-6} \cmidrule(lr){7-8}
        & Correct & Wrong & 
        Naturalness $\uparrow$ & Realism $\uparrow$  & Smoothness$\uparrow$  & Floating [mm] $\downarrow$ & Jerk [mm/s$^3$] $\downarrow$\\
        \midrule
        Phys-GT & - &  - &3.43 $\pm$ 1.11 & 3.53 $\pm$ 0.99 & 3.42 $\pm$  0.93&  17.0 &  1.1\\
        \midrule
        CLoSD \cite{tevet2024closd} & \textbf{61.6}\% & \textbf{8.6}\% & 2.86 $\pm$  1.06& 3.06 $\pm$ 1.00& 2.86 $\pm$ 0.97 & 20.59 &  3.4 \\
        MaskedMimic \cite{tessler2024masked} & 42.9\% & 16.6\% &2.82 $\pm$  1.07& 2.93 $\pm$ 1.05& 2.78 $\pm$ 1.06 & \textbf{14.82} &  4.2 \\
        
        \midrule
        UniPhys (Ours) & 56.3\% & 14.2\% & \textbf{3.23 $\pm$ 1.19} & \textbf{3.28 $\pm$  1.13} & \textbf{3.15 $\pm$ 1.01} & 16.6 &  \textbf{1.2} \\
        \bottomrule
    \end{tabular}
    \caption{Evaluation on the text-driven controller and comparison with baselines. Phys-GT refers to the physics-based motions that are tracked from the MoCap dataset.}
    \label{tab:baselines}
\end{table*}

\subsection{Text-Driven Control Evaluation.}
\myparagraph{Evaluation setup.} 
In the absence of reliable automatic metrics for physics-based text-driven control, we conduct extensive user studies to evaluate the naturalness and expressiveness of our text-driven control policy. To assess semantic fidelity, we adopt the user study design from SuperPADL~\cite{juravsky2024superpadl}, where raters view a generated motion together with four candidate captions: one ground truth and three distractors. Options also include `Nothing applies' and `Multiple apply' to address ambiguous or similar captions. For motion quality, raters evaluate each motion on naturalness, realism, and smoothness on a scale of 1 to 5, with higher being better. Each motion is assessed by three raters to ensure reliability. For quantitative evaluation and baseline comparison, we randomly sample 150 captions from the BABEL \cite{BABEL:CVPR:2021} validation set, covering various skills like walking, exercise, and object interaction, with all motions starting from a standardized neutral standing pose for a fair comparison.

\myparagraph{Results.} Table~\ref{tab:baselines} presents the quantitative comparison against baselines, including user studies and automatic motion quality metrics. Our generated motions consistently outperform baselines in naturalness and smoothness, as evidenced by user study scores and the motion jerk metric. Compared to MaskedMimic, our method achieves superior semantic fidelity and motion quality. While CLoSD's kinematics-based motion generation is more expressive than physics-based text-driven controllers, its planning-then-tracking paradigm results in jittery motion and more foot-floating artifacts. 
Our approach provides responsive text-driven control with smooth transitions between skills. More qualitative results are available in the Supp. Mat.

\begin{figure}
    \centering
    \includegraphics[width=1.0\linewidth]{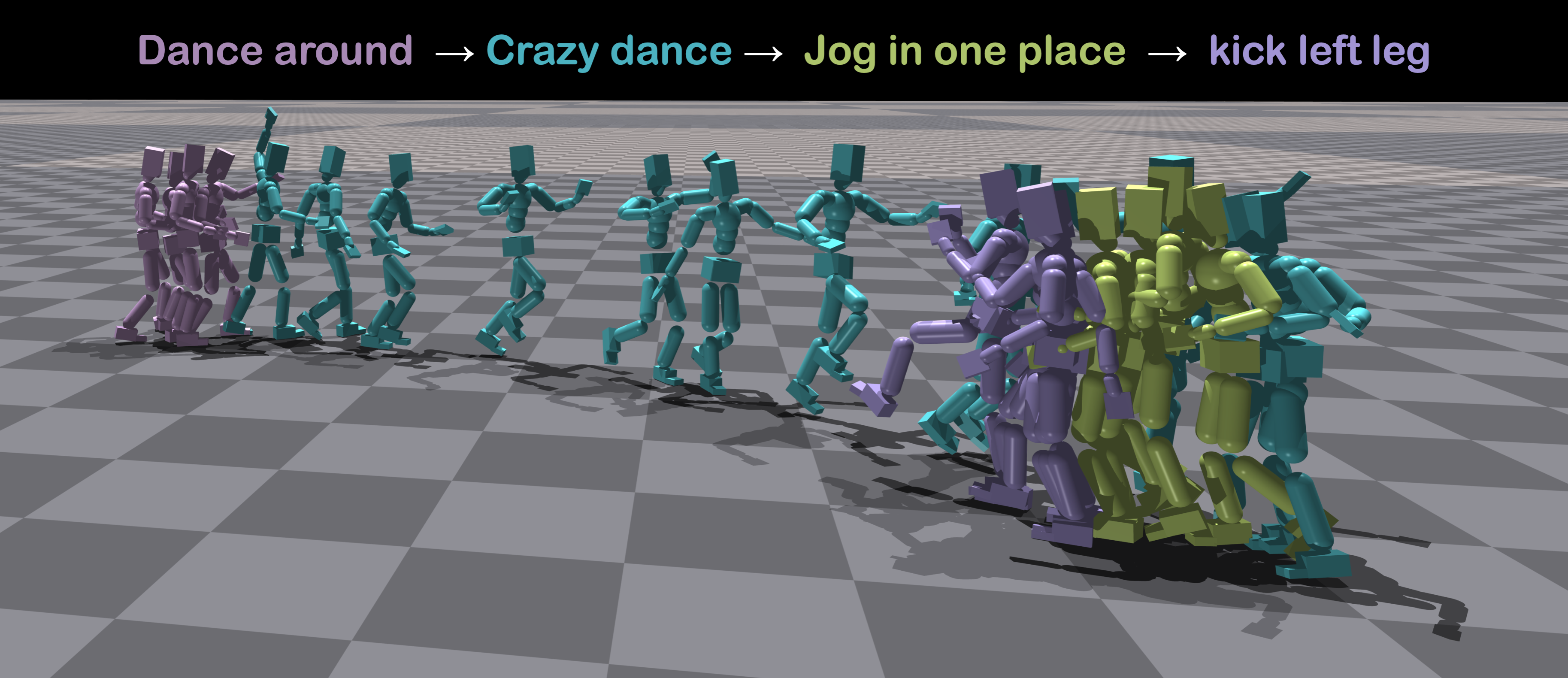}
    \caption{Expressive text-driven control with smooth transition between skills.}
    \label{fig:enter-label}
\vspace{-0.2cm}
    
\end{figure}


\subsection{Goal Reaching.} 

\myparagraph{Evaluation setup.} We randomly assign target location coordinates as goals for the pelvis. The target height is fixed at 0.9m following CLoSD's setup as CLoSD does not support height control. A task is successful if the character is within 0.3m of the goal. We evaluate goal-reaching in two scenarios: (1) near goals between 1–2 meters and (2) far goals between 2–6 meters. Additionally, we assess performance in different motion styles, such as `walking' and the more challenging `jogging'. Success rates for each setup are reported in Table~\ref{tab:multi-task-control} along with motion quality assessed via floating and jerk.

\myparagraph{Results.} Both CLoSD and MaskedMimic constrain goal-reaching targets by always conditioning on target positions within a two-second horizon, making them less effective for distant goals. CLoSD attempts to overcome this by setting intermediate goals heuristically, however, doing so often causes motion discontinuities. Furthermore, CLoSD struggles with the `jogging' style due to its RL-based tracking controller failing to execute the planned kinematic motions. With a planning-then-tracking approach, CLoSD’s results frequently contain high jitter, excessive jerk, and foot-floating.
In contrast, our method effectively generalizes to goals at any distance without heuristic goal setting. The motion generated by our end-to-end behavior controller is significantly smoother and more natural than baseline methods. Please check the Supp. Mat. for qualitative results.



\subsection{Velocity Control.} We randomly assign target velocities with random directions and speeds ranging from 0 to 2 m/s, updating every 500 steps, and evaluate the policy over 20 target velocity transitions. We allow a 120-step (4-second) transition period before assessing velocity tracking error. Since CLoSD lacks velocity control, we compare only with MaskedMimic. Our method achieves similar tracking errors but produces significantly smoother, more natural motion (see supplementary video). In contrast, MaskedMimic shows abrupt accelerations during transitions, resulting in high motion jerks.

\subsection{Dynamic Obstacle Avoidance.} We randomly spawn a spherical obstacle approximately 2 meters away from the character, moving continuously toward it. We use a sphere-shaped obstacle to simplify SDF calculations. The task is successful if the character actively increases its distance from the obstacle to at least 3 meters while avoiding collisions and maintaining balance. By predicting and dynamically re-planning, our method effectively adapts to obstacles from various directions. In 50 episodes with different obstacle approaches, we achieve a 94\% success rate. Failures mainly occur when the obstacle approaches from the front of the character, causing it to step backward and lose balance.


\begin{table*}[]
    \centering
    \footnotesize
    \begin{tabular}{lccccc|ccc}
    \toprule
    & \multicolumn{5}{c}{\textbf{Goal Reaching}}
    & \multicolumn{3}{c}{\textbf{Velocity Control}}
    \\
    \cmidrule(lr){2-6} \cmidrule(lr){7-9} 
    \multirow{2}{*}
    & \multicolumn{3}{c}{\textbf{Success Rate (SR)}} & \textbf{Floating $\downarrow$} & \textbf{Jerk $\downarrow$ }&  \textbf{Error} & \textbf{Floating $\downarrow$} & \textbf{Jerk $\downarrow$}  
    \\
    \cmidrule(lr){2-4}
         & \textit{[walk]-near} & \textit{[walk]-far} & \textit{[jog]} & [mm] & [mm/s$^3$] & [m/s] & [mm] & [mm/s$^3$] 
         \\
         \midrule

    CLoSD \cite{tevet2024closd} & 1.0 & 1.0 & 0.64 & 25.19 & 3.8 & - & - & -  \\
    MaskedMimic \cite{tessler2024masked} & 0.87 & 0.0 & 0.81$^*$ & 16.68 & 2.5  & 0.09 & 18.69 & 5.0 \\
    \midrule
    UniPhys (Ours) & 1.0 & 0.95 & 0.85 & 16.78 & 1.5 & 0.07 & 19.04 & 2.0  \\
    \bottomrule
    \end{tabular}
    \caption{Evaluation of the multi-task control and comparison with baselines. $^*$ For \textit{jog-to-goal} task, as MaskedMimic cannot handle far-away goals, we set close goals (within 2m) so that it can still reach, while for CLoSD and our UniPhys, we set the goals that are within 6m.}
    \label{tab:multi-task-control}
\vspace{-0.2cm}
    
\end{table*}

\subsection{Ablation study}
We systematically evaluate the impact of various design choices in our framework on control policy robustness and inference-time efficiency, as shown in Table~\ref{tab:ablation-robustness}. For robustness assessment, we randomly generate episodes for unconditional behavior synthesis, ending an episode when the character falls or after 3000 steps (100-second motion). We conduct 150 episode rollouts and report the mean and maximum episode lengths and frames per second (FPS). FPS is measured when using DDIM sampling with 5 denoising steps on an Nvidia A100 GPU.

\myparagraph{Latent Action Representation.} Learning upon the latent action representation significantly enhances efficient and robust policy learning compared to directly learning from high-dimensional raw action space.

\myparagraph{Stabilization Trick.} Regardless of the action space used, the stabilization trick consistently stabilizes policy rollout by reducing compounding errors.

\myparagraph{Different Denoising Schedules.} An autoregressive denoising schedule, combined with the stabilization trick, yields the most robust policy but is the least efficient compared to full-sequence and gradual denoising schedules. The gradual denoising schedule offers a good balance between robustness and efficiency.


\begin{table}[]
    \centering
    \footnotesize
    \begin{tabular}{ccc|ccc}
    \toprule
        Denoising &Latent &  \multirow{2}{*}{Stabilization}& \multicolumn{2}{c}{Episode Length} & \multirow{2}{*}{FPS} \\
        \cmidrule(lr){4-5}
        schedule &action &  & Mean & Max  & \\
        \midrule
    Full-sequence & \ding{51} & \ding{51} & 231.6 & 1197 & \textbf{23.0} \\
    Gradual &\ding{51} &  \ding{51}  & 1817.6& \textbf{3000} & 18.0 \\
    \midrule
    Autoregressive &\ding{55} &  \ding{55}  & 58.0 & 59 & 9.5 \\
    Autoregressive &\ding{55} &  \ding{51}  & 238.2 & 717 & 9.5 \\
    Autoregressive &\ding{51} &  \ding{55}  & 148.2 & 230 & 9.5 \\
    Autoregressive &\ding{51} &  \ding{51}  & \textbf{2320.3}& \textbf{3000} & 9.5\\

    \bottomrule
    \end{tabular}
    \caption{Ablation study on the effect of different design choices on policy robustness and efficiency.}
    \label{tab:ablation-robustness}
\vspace{-0.2cm}
    
\end{table}




%% file: 10_conclusion.tex
\section{Conclusion}
\label{sec:conclusion}
We introduce UniPhys, a unified diffusion-based planner and controller for physics-based character control, bridging the gap in previous works that separate high-level planning and low-level control. Our end-to-end framework enhances motion coherence and provides flexibility to handle diverse or unseen control signals using high-level text instructions and task-specific guidance. We improve motion stability within the behavior cloning framework by effectively reducing compounding errors through the Diffusion Forcing training paradigm. UniPhys not only expands possibilities for downstream tasks but also provides a foundation for extending character control to more complex tasks such as dexterous hand manipulation. Using a compact latent action representation, our method is well-suited for higher-dimensional action space predictions, paving the way for future research in physics-based character control.


%% file: 12_appendix.tex
\appendix

\twocolumn[
\begin{center}
\Large{\bf UniPhys: Unified Planner and Controller with Diffusion for Flexible Physics-Based Character Control \\ \vspace{0.5em} **Appendix**}
\end{center}
\vspace{1em}
]

\counterwithin{table}{section}
\counterwithin{figure}{section}
\setcounter{page}{1}

We provide comprehensive qualitative results on diverse tasks and qualitative comparisons with baselines in the supplementary HTML file. We strongly encourage readers to check them by clicking \href{run:supp_video.html}{here}. 

Additionally, we include further implementation details for both training and inference in Sec.\ref{supp:implementation-details}. In Sec.\ref{supp:application}, we detail the loss design for tasks utilizing loss-based guided sampling. Sec.~\ref{supp:user-study} presents the user study design, interface, and complete results on text-to-motion alignment evaluation. 
Finally, we discuss the limitations of our approach and potential directions for future work.


\section{Implementation Details}
\label{supp:implementation-details}
\myparagraph{Architecture.} The diffusion model is build with a 12-layer causal transformer decoder with a hidden size of 768. The input is a sequence with 32 frames, and the per-frame input feature includes the 32-dim latent action embedding and the 366-dim state representation.

\myparagraph{Training details.} During training, we divide motion sequences into 32-frame clips with a stride of 8. If a clip contains multiple text annotations, we randomly select one for training. To improve transition smoothness between different skills, we preprocess the annotations by removing "transition to" and assigning the annotation of transition-phase motion to the target motion.

We train the model with a batch size of 1024, a learning rate of $1.5 \times 10^{-4}$, 10k warm-up steps, and cosine learning rate decay. The model undergoes training with 50 denoising steps, taking approximately 10 GPU days on a single RTX A100 over 15k epochs. Despite only a minor decrease in loss as training goes on, we still observe continuous improvements in policy stability and motion-semantic fidelity.

\myparagraph{Inference details.} At inference time, we use DDIM sampling with 5 steps and apply the stabilization trick across all applications.

\textbf{(a) Text-Driven Control Policy:} We empirically find that a small stabilization noise level (1, 2, or 3) is sufficient for achieving stable long-horizon control, whereas increasing it further to 5 degrades stability. Therefore, we use a stabilization noise level of 3 for all text-driven control experiments.

\textbf{(b) Loss-Based Guided Applications:} For challenging tasks that utilize loss-based guidance, we observe that increasing the stabilization noise level helps stabilize the guided denoising process. Intuitively, a strong task-specific guidance signal may cause the denoised states to drift slightly out of distribution, and a higher stabilization noise level mitigates this effect.

Moreover, we employ Monte Carlo guidance by estimating the gradient from multiple samples to reduce gradient variance and stabilize the guided optimization process. Without Monte Carlo guidance, the optimization tends to be unstable, resulting in a low task success rate.

We analyze the effect of Monte Carlo guidance on the goal-reaching task in Table~\ref{tab:ablate-MCG}. With just 2 Monte Carlo samples, the success rate significantly improves from 26\% to 82\%. Increasing the number of samples to 5 further enhances performance, though at the cost of slightly reduced planning efficiency.

\begin{table}[]
    \centering
    \footnotesize
    \begin{tabular}{lccc}
    \toprule
        MC Samples & N=1 & N=3 & N=5 \\
        \midrule
        Succ. Rate & 26\% & 82\% & 98\% \\
        FPS & 9.2 & 8.9 & 8.7 \\
        \bottomrule
    \end{tabular}
    \caption{Ablation on the effect of Monte-Carlo Guidance (MCG) on loss-based guided sampling for goal reaching task.}
    \label{tab:ablate-MCG}
\end{table}


\section{Loss-guided sampling design details}
\label{supp:application}

\myparagraph{Goal reaching.} 
To facilitate this goal reaching process, we design a loss function that encourages the predicted joint position to be close to the target goal. Furthermore, to expedite goal achievement, we incorporate an orientation loss that encourages the character to orient itself toward the goal. Specifically, the loss function is defined as follows:
\begin{align}
    \mathcal{G}(\hat{\mathbf{X}}) &= \sum_{i=t+1}^{t+H}(w_1 * |\hat{\mathbf{p}}_i - \mathbf{p}^g| \nonumber \\
    &+ w_2 * (1 - cos<\hat{\mathbf{\phi}}_i, \mathbf{p}^g-\hat{\mathbf{p}}_i>))
\end{align}
where $\mathbf{p}$ and $\mathbf{p}^g$ are the joint position and goal position, respectively, and $\mathbf{\phi}$ is the character root orientation, and $w_1, w_2$ adjust the strength of position guidance and orientation guidance.

\myparagraph{Velocity Control.} For velocity control, we apply losses the speed magnitude, the steering direction and also the orientation direction to align the character's oriention with the target velocity. The loss function is formulated as follows:
\begin{align}
    \mathcal{G}(\hat{\mathbf{X}}) &= \sum_{i=t+1}^{t+H} (w_1 \left\| \|\mathbf{v}_t\| - \|\mathbf{v}^g\| \right\|^2  \nonumber \\
    &+ w_2 \left(1 - \cos{\theta_v} \right) 
    + w_3 \left(1 - \cos{\theta_o} \right)),
\end{align}
where $\mathbf{v}_t, \mathbf{v}^g$ is the predicted velocity and the target velocity respectively, and $\theta_v$ is the angle between $\mathbf{v}_t$ and $\mathbf{v}^g$, and $\theta_o$ is the angle between the character's orientation and $\mathbf{v}^g$, ensuring the agent faces the movement direction, and $w_1, w_2, w_3$ balances the guidance strength of each term.

\myparagraph{Dynamic Obstacle Avoidance.} We employ a smooth SDF-based loss with softplus smoothing, and for SDF computing simplicity, we adopt for the sphere-like obstacle, and the guidance loss is designed as follows,
\begin{align}
    \mathcal{G}(\hat{\mathbf{X}}) &= \sum_{i=t+1}^{t+H}\log(1+e^{-(d_i-r-1)})
\end{align}
where $d_i$ is the distance between the character's root and obstacle's center in XY plane, and $r$ is the radius of the obstacle.

\section{User study interface and more results}
\label{supp:user-study}
\begin{figure*}[ht]
    \centering
    \includegraphics[width=1\linewidth]{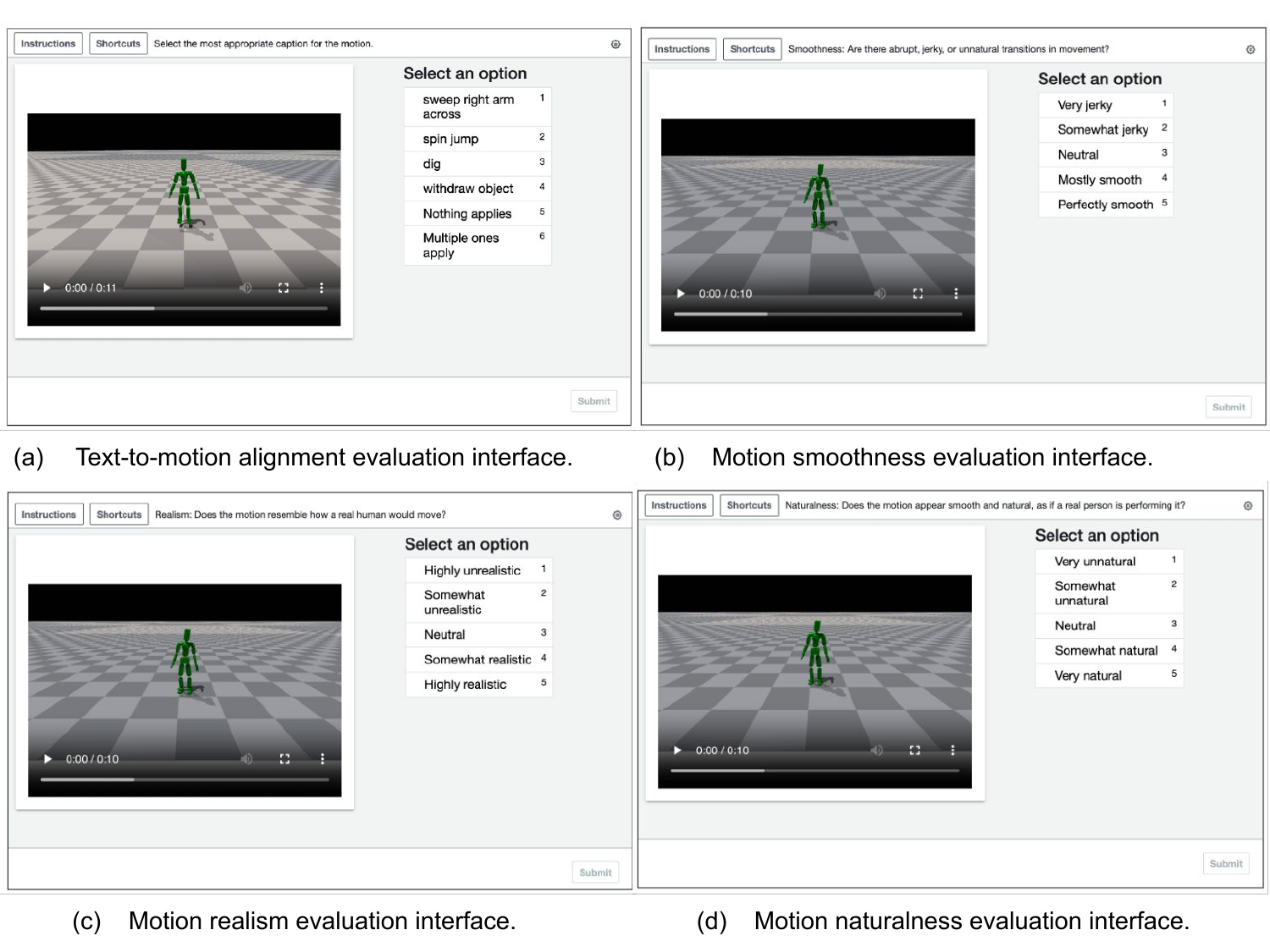}
    \caption{User study interface on the Amazon Mechanical Turk (AMT).}
    \label{fig:user-study}
\end{figure*}

We conduct two user studies on Amazon Mechanical Turk to evaluate motion semantic fidelity and motion quality separately.

For motion semantic fidelity, we follow the evaluation protocol from SuperPADL~\cite{juravsky2024superpadl}. Raters are presented with four options per motion (three distractors and one ground truth) and can also select "Nothing applies" or "Multiple ones apply" to account for annotation ambiguity. To ensure fair comparisons between our method and baselines, we use the same text prompts for motion generation and provide identical answer choices for each motion.

For motion quality, we ask raters to assess naturalness, smoothness, and realism to make the results more interpretable. All motions are initialized with a standing pose, and we ask 3 independent raters to rate each motion. The user study interface is shown in Fig.~\ref{fig:user-study}, and in Table~\ref{table:user-study}, we present the complete user study results on the text-to-motion alignment evaluation.

\begin{table}[]
    \centering
    \footnotesize
    \begin{tabular}{c|ccc}
    \toprule
       User Response  &  Ours & CLoSD & MM\\
    \midrule
       Correct   & 56.3\% &  61.6\%& 42.9\%\\
       Wrong  & 14.2\%  &  8.6\% & 16.6\%\\
       Nothing applies  & 23.8\% &  21.7\% & 35.1\%\\
       Multi apply  & 5.6\% &  7.9\% & 5.3\%\\
    \midrule
        Any Correct  & 92.7\% & 94.5\% & 79.3\%\\
        Majority Correct  & 52.0\% & 65.3\% & 34.6\%\\
        All Correct  & 24.0\% & 25.1\% & 14.6\% \\
       \bottomrule
    \end{tabular}
    \caption{Complete user study results on the text-to-motion semantic alignment evaluation.}
    \label{table:user-study}
\end{table}


\section{Limitations and future work}
Inference inefficiency is a common limitation of diffusion-based frameworks, making our method less efficient than RL-based policies. For text-driven control, our framework operates at approximately 10 FPS with autoregressive denoising and 18 FPS with gradual denoising. However, improving inference efficiency was not the primary focus of this work. Recent advancements in diffusion-based kinematic motion generation~\cite{Zhao:DART:2025, tevet2024closd} have demonstrated real-time interactive motion generation. We believe that further optimizations in diffusion model inference could enable our framework to be applied to high-frequency, real-time control tasks.

While our model demonstrates robust control, balance loss still occurs, particularly during highly dynamic actions or due to poor timing in changing text instructions, leading to skill transition failures and falls. Completely avoiding falls is unrealistic due to the inherent challenges in bipedal control tasks. Moving forward, we plan to incorporate a fall recovery skill by collecting expert demonstrations on getting up from the ground and leveraging an RL policy specifically trained for this task to enhance the expert demonstration data collection.

Another interesting capability for physics-based character control is traversing different terrains, which is crucial for real-world applications, such as robotics. Due to the lack of terrain-specific data, achieving this under a behavior cloning framework is not immediately feasible. However, reinforcement learning-based policies can serve as a valuable data generator for unseen scenarios, making it possible to explore the potential of behavior cloning in this context.

Lastly, our current approach does not incorporate dexterous hand control for the character, limiting its application in tasks like human-object interaction. However, our framework can be extended to full-body character control, including hand dexterity.
